\documentclass{natureprintstyle}
\usepackage{epsfig,caption}
\usepackage{color}
\usepackage{bm}
\usepackage{graphicx}
\usepackage{longtable}
\usepackage{amsmath}
\usepackage{amssymb}
\usepackage{rotating,xcolor}
\usepackage{hyperref}
\usepackage{subcaption}
\usepackage[]{chapterbib}
\usepackage{enumitem}
\begin{document}

\newcommand{\actaa}{Acta Astron.}   
\newcommand{\araa}{Annu. Rev. Astron. Astrophys.}   
\newcommand{\areps}{Annu. Rev. Earth Planet. Sci.} 
\newcommand{\aar}{Astron. Astrophys. Rev.} 
\newcommand{\ab}{Astrobiol.}    
\newcommand{\aj}{Astron. J.}   
\newcommand{\ac}{Astron. Comput.} 
\newcommand{\apart}{Astropart. Phys.} 
\newcommand{\apj}{Astrophys. J.}   
\newcommand{\apjl}{Astrophys. J. Lett.}   
\newcommand{\apjs}{Astrophys. J. Suppl. Ser.}   
\newcommand{\ao}{Appl. Opt.}   
\newcommand{\apss}{Astrophys. Space Sci.}   
\newcommand{\aap}{Astron. Astrophys.}   
\newcommand{\aapr}{Astron. Astrophys. Rev.}   
\newcommand{\aaps}{Astron. Astrophys. Suppl.}   
\newcommand{\baas}{Bull. Am. Astron. Soc.}   
\newcommand{\caa}{Chin. Astron. Astrophys.}   
\newcommand{\cjaa}{Chin. J. Astron. Astrophys.}   
\newcommand{\cqg}{Class. Quantum Gravity}    
\newcommand{\epsl}{Earth Planet. Sci. Lett.}    
\newcommand{\frass}{Front. Astron. Space Sci.}    
\newcommand{\gal}{Galaxies}    
\newcommand{\gca}{Geochim. Cosmochim. Acta}   
\newcommand{\grl}{Geophys. Res. Lett.}   
\newcommand{\icarus}{Icarus}   
\newcommand{\jatis}{J. Astron. Telesc. Instrum. Syst.}  
\newcommand{\jcap}{J. Cosmol. Astropart. Phys.}   
\newcommand{\jgr}{J. Geophys. Res.}   
\newcommand{\jgrp}{J. Geophys. Res.: Planets}    
\newcommand{\jqsrt}{J. Quant. Spectrosc. Radiat. Transf.} 
\newcommand{\lrca}{Living Rev. Comput. Astrophys.}    
\newcommand{\lrr}{Living Rev. Relativ.}    
\newcommand{\lrsp}{Living Rev. Sol. Phys.}    
\newcommand{\memsai}{Mem. Soc. Astron. Italiana}   
\newcommand{\mnras}{Mon. Not. R. Astron. Soc.}   
\newcommand{\nat}{Nature} 
\newcommand{\nastro}{Nat. Astron.} 
\newcommand{\ncomms}{Nat. Commun.} 
\newcommand{\nphys}{Nat. Phys.} 
\newcommand{\na}{New Astron.}   
\newcommand{\nar}{New Astron. Rev.}   
\newcommand{\physrep}{Phys. Rep.}   
\newcommand{\pra}{Phys. Rev. A}   
\newcommand{\prb}{Phys. Rev. B}   
\newcommand{\prc}{Phys. Rev. C}   
\newcommand{\prd}{Phys. Rev. D}   
\newcommand{\pre}{Phys. Rev. E}   
\newcommand{\prl}{Phys. Rev. Lett.}   
\newcommand{\psj}{Planet. Sci. J.}   
\newcommand{\planss}{Planet. Space Sci.}   
\newcommand{\pnas}{Proc. Natl Acad. Sci. USA}   
\newcommand{\procspie}{Proc. SPIE}   
\newcommand{\pasa}{Publ. Astron. Soc. Aust.}   
\newcommand{\pasj}{Publ. Astron. Soc. Jpn}   
\newcommand{\pasp}{Publ. Astron. Soc. Pac.}   
\newcommand{\raa}{Res. Astron. Astrophys.} 
\newcommand{\rmxaa}{Rev. Mexicana Astron. Astrofis.}   
\newcommand{\sci}{Science} 
\newcommand{\sciadv}{Sci. Adv.} 
\newcommand{\solphys}{Sol. Phys.}   
\newcommand{\sovast}{Soviet Astron.}   
\newcommand{\ssr}{Space Sci. Rev.}   
\newcommand{\uni}{Universe} 
\title{A Tilted Dark Halo Origin of the Galactic Disk Warp and Flare}

\author{Jiwon Jesse Han, Charlie Conroy, and Lars Hernquist \\ {\small Center for Astrophysics $|$ Harvard \& Smithsonian,  Cambridge, MA, USA}}

\maketitle

\begin{abstract}

  The outer disk of the Milky Way Galaxy is warped and flared. Several mechanisms have been proposed to explain these phenomena, but none have quantitatively reproduced both features. Recent work has demonstrated that the Galactic stellar halo is tilted with respect to the disk plane, suggesting that at least some component of the dark matter halo may also be tilted. Here we show that a dark halo tilted in the same direction as the stellar halo can induce a warp and flare in the Galactic disk at the same amplitude and orientation as the data. In our model the warp is visible in both the gas and stars of all ages, which is consistent with the breadth of observational tracers of the warp. These results, in combination with data in the stellar halo, provide compelling evidence that our Galaxy is embedded in a tilted dark matter halo. This misalignment of the dark halo and the disk holds clue to the formation history of the Galaxy, and represents the next step in the dynamical modeling of the Galactic potential.

\end{abstract}
\vspace{.2cm}

\begin{figure}[t]
    \centering
    \includegraphics[width=0.48\textwidth]{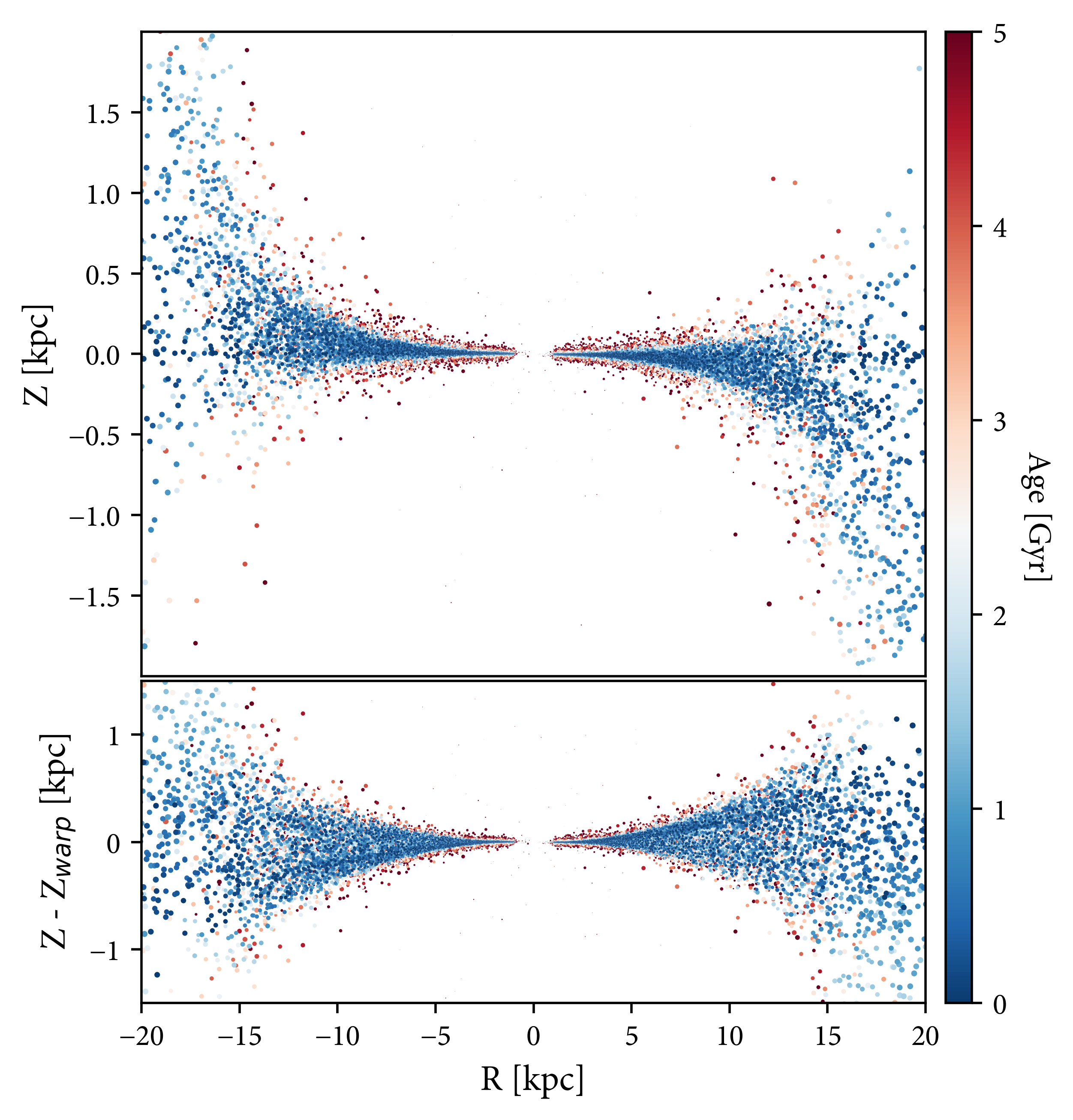}
    \caption{Present-day distribution of simulated stars in Galactocentric cylindrical coordinates. Negative R indicates azimuthal angles that are within $90^\circ$ of the Northern warp, and positive R indicates azimuthal angles that are within $90^\circ$ of the Southern warp. The top panels show the warp, and the bottom panels show the vertical deviation from the average warp. The vertical deviation systematically increases toward the outer Galaxy, demonstrating the flare of the disk. Points that are in the outer Galaxy are plotted with larger markers. Particles colored by stellar age clearly demonstrate a warp at all ages, and most strongly at the youngest population.}
    \label{fig:model age}
\end{figure}

We construct a model of the gravitational potential of the Galaxy in which 30\% of the dark halo mass is comprised of a triaxial distribution that is tilted $25^\circ$ with respect to the disk plane. The potential also includes bulge and disk components, where the latter increases in time to mimic the growth of the Galactic disk with time.  We calculate the orbits of collisionless (stars) and collisional (gas) particles over 5 Gyr in this potential. In Figure \ref{fig:model age} we show the present-day distribution of stars in the simulation. In the top panel we plot the Galactocentric vertical height ($Z$) and signed cylindrical radius ($R$), clearly revealing an S-shaped \textit{warp} of the disk. Negative $R$ indicates the Galactic quadrants that are within $\pm90^\circ$ of the positive peak (the ``Northern'' warp) and positive $R$ indicates Galactic quadrants that are within $\pm90^\circ$ of the negative peak (the ``Southern'' warp). In the bottom panels, we plot the vertical deviation ($Z-Z_{warp}$) of tracer particles from the mean warp. The average vertical deviation grows markedly towards outer radii, demonstrating the \textit{flare} of the disk. Particles are color-coded by age, which shows that stars of all ages exhibit a warp and flare. The warp is most pronounced for the youngest stars, consistent with observations\cite{chen19,wang20}.

\begin{figure*}[t]
    \centering
    \includegraphics[width=0.9\textwidth]{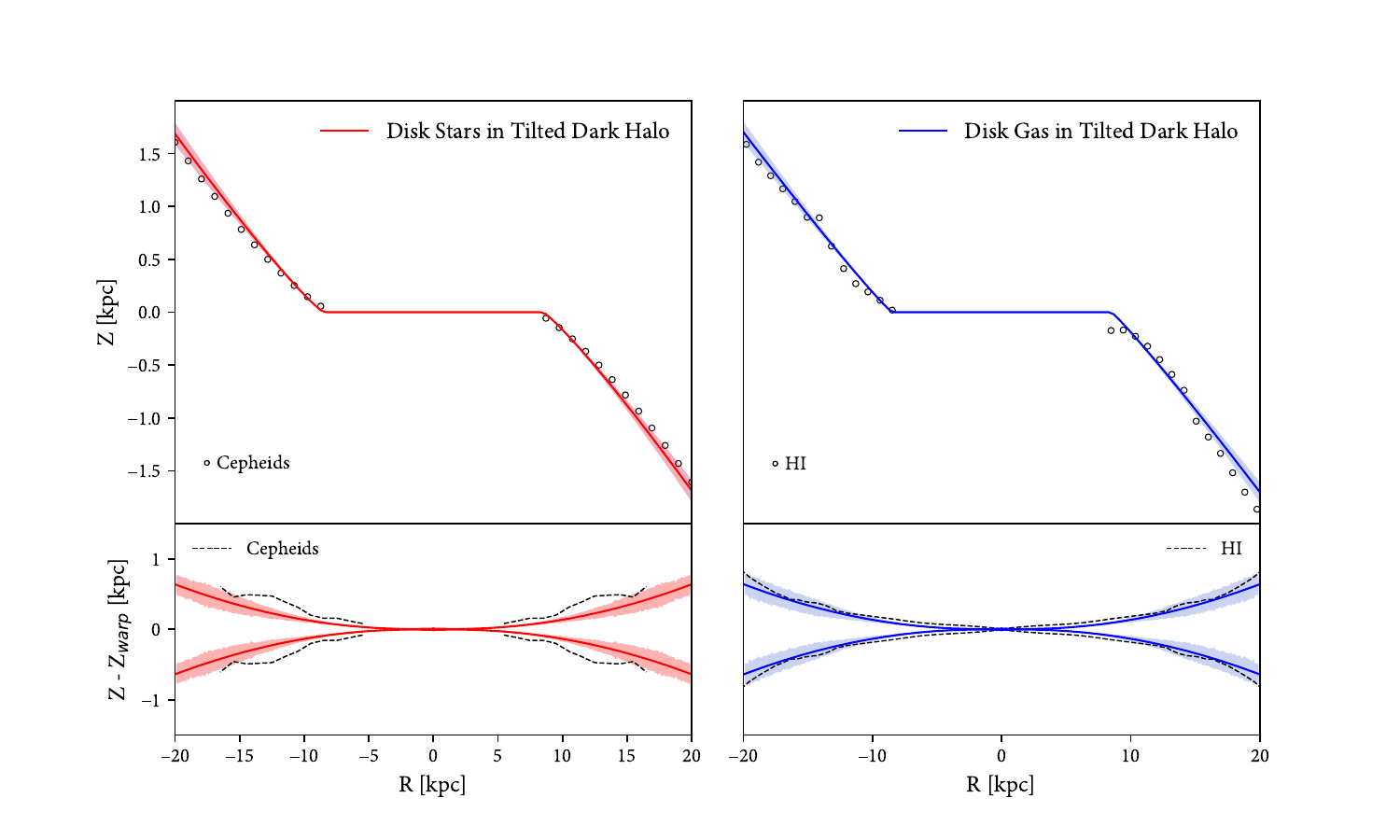}
    \caption{Comparison of the simulation to the observed warp and flare. The top panels show the warp, and bottom panels show the flare. The observed points (open circle) are extracted directly from the cited papers\cite{chen19,levine06,Kalberla07}. (a) In the top panel, the observed warp of Cepheids\cite{chen19} is plotted in open circles, and the maximum-likelihood fit of stellar warp in the simulation is plotted in red. We fix the onset radius of the warp to the observed value. Shaded regions indicate $1\sigma$ uncertainty of the fit. In the bottom panel, the observed scale height of Cepheids are plotted in black dashed lines, and the scale height of the simulated stars are plotted in red. (b) In the top panel, the observed warp of HI\cite{levine06,Kalberla07} is plotted in open circles, and the maximum-likelihood fit of the gaseous warp in the simulation is plotted in blue. In the bottom panel, the observed scale height of HI is plotted in black dashed lines, and the scale height of the simulated gas are plotted in blue.}
    \label{fig:lit}
\end{figure*}

In Figure \ref{fig:lit} we compare the warp and flare of the simulated stars (left panel) and gas (right panel) to observations. For conciseness of comparison, we focus on the HI\cite{levine06, Kalberla07} gas and Cepheid\cite{chen19} stars. We note that a similar warp has also been observed for ionized gas\cite{Dame11}, molecular gas\cite{Wouterloot90}, dust\cite{drimmel01}, stars\cite{Djorgovski89, yusifov03, momany06, wang20}, and star clusters\cite{cantat-gaudin20}. In the top panels we plot the observed warp in open circles, and a maximum likelihood fit to the simulation in solid colored lines, with $1 \sigma$ uncertainty of the fit shaded. The onset radius $R_w$ of the fit is fixed to the observed $R_w$ values.  The flare is measured by fitting an exponential scale height to the vertical deviation from the warp at each radius. In both stars and gas, the simulated warp and flare quantitatively match the observations. Additionally, the tilted halo simulation produces a population of stars on circular orbits that reach high-latitude ($|Z| > 2\text{ kpc}$) on either side of the warp extrema. This population of stars is reminiscent of known stellar overdensities towards the Galactic anticenter\cite{newberg02,martin04,momany06,naidu20}.  

The parameters chosen for the simulation were not tuned to match the Galactic warp or flare. The two key parameters affected the predicted warp and flare are: (1) the shape and extent of the tilted halo, and (2) the scale radius of newly-formed stars. For the former, we adopt the scale radius, triaxiality, and tilt angle of the halo directly from the shape of the accreted stellar halo\cite{han22b}. For the latter, we adopt the scale radius of the molecular disk of the Galaxy\cite{heyer15}. Details of the simulation setup and variations to the adopted parameters are given in the Supplementary Material.

\begin{figure*}[t!]
    \centering
    \includegraphics[width=\textwidth]{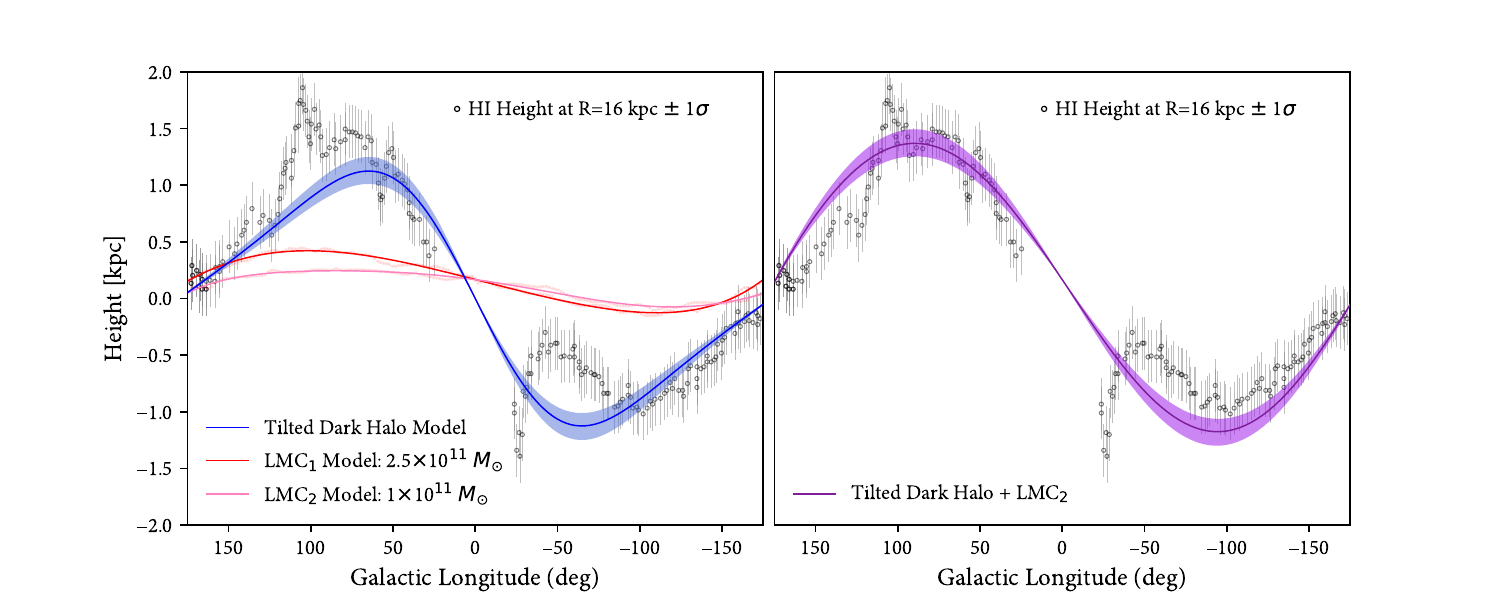}
    \caption{The amplitude and orientation of the disk warp at $R=16\text{ kpc}$. In the left panel we plot the simulated warp from the tilted dark halo in blue, the LMC models\cite{laporte18} in red, and the HI warp in open circles\cite{levine06}.  Shaded regions indicate $1\sigma$ uncertainty of the fit. The individual data points from the LMC simulation are marked in faint dots, and a polynomial fit to the points are drawn as solid lines. In the right panel, we show the sum of the warps induced by a tilted halo and the LMC.}
    \label{fig:r_16}
\end{figure*}

The large-scale warp of the Galactic disk has been known for over half a century\cite{burke57,kerr57}, and there is a commensurately rich history of warp and flare models\cite{hunter69,toomre83,olling98,olling00}. For example, studies have investigated the warp as a result of perturbed bending modes\cite{sparke88}, misaligned angular momenta of the halo and the disk\cite{debattista99, jiang99}, repeated impact from the Sagittarius dwarf galaxy\cite{poggio21}, misaligned gas accretion\cite{ostriker89}, or quadrupolar torque from a tumbling triaxial halo\cite{dubinski09}. However, previous studies have been unable to quantitatively reproduce the warp (and simultaneously the flare) of the Galactic disk. Among the recently investigated models is the tidal influence of the Large Magellanic Cloud (LMC)\cite{laporte18}. In Figure \ref{fig:r_16} we plot the disk warp at $R=16\text{ kpc}$ produced by the tilted halo model and an LMC model as a function of Galactic longitude. The LMC was modeled as a live halo on its first infall into a live Milky Way halo and disk\cite{laporte18}. The HI data\cite{levine06} at this radius is marked in open circles. Even for the highest LMC mass---which is 80\% higher than other models\cite{erkal19}---the warp amplitude is less than a third of the observed amplitude. In the right panel, we sum the amplitudes of the warp from the LMC and the tilted halo models. The combined model matches the observations better than the tilted halo alone, although we caution that this simple summation does not capture the possible coupling between the effect of the LMC and the tilted halo.

Tilted dark halos are common in galaxy simulations that include baryonic physics\cite{Prada19,emami21a}, and several independent lines of evidence have pointed toward a tilted Galactic dark halo\cite{debattista13,shao21,han22a}. Furthermore, Han et al. (in prep) show that such tilted dark halos are long-lived and can warp Galactic disks in Illustris TNG50\cite{pillepich19, nelson19}. In a realistic model we can expect the tilt angle to change with time, for example due to interaction between the (growing) disk and the halo. While the tilt of the halo at earlier epochs can leave interesting observational signatures in old disk stars, the main point of this paper is to demonstrate that the young disk (and gas) is responsive to the tilt on $<1$ Gyr timescales, and on such timescales the tilt is constrained observationally to be $\sim25^\circ$. A plausible origin of a tilted dark halo in the Galaxy is a major merger\cite{belokurov18,helmi18} $8-10\text{ Gyr}$ ago\cite{gallart19,bonaca20} that deposited a significant fraction of dark matter on an eccentric, tilted orbit\cite{naidu21,han22a}. Dynamical models of the Galaxy often assume a spherical dark halo at all scales, or a flattened halo that is aligned with the disk. A tilted dark halo at $\sim30\text{ kpc}$ would have novel applications for Galactic dynamics. For example, a tilted and triaxial halo influences the shape of the stellar halo\cite{han22a}, and can affect orbit reconstruction of stellar streams\cite{Lilleengen23}. Furthermore, the tilt and triaxiality of the inner halo ($\sim0.1 R_{\text{virial}}$) encodes information about the self-interacting properties of dark matter\cite{peter13} that is unique compared to larger-scale probes such as galaxy clusters or large scale structure. Future work will investigate the imprint of dark matter self-interaction through the tilt and triaxiality of the dark halo at $30\text{ kpc}$. In addition, a global tilt in the dark halo implies an anisotropic velocity distribution of the dark matter particles. The resulting asymmetric velocity distribution should affect ground-based dark matter detection experiments\cite{evans19}.

Finally, while we have focused on demonstrating that the warp and flare are likely manifestations of a tilted dark halo, we can also reverse the argument: precise measurements of the warp may further constrain the tilt of the dark halo. We have intentionally avoided ``fitting'' a dark halo to match the warp, but there is much to explore in allowing for more flexible halo models. For example, the warp is sensitive to the tilt angle of the dark halo, and the fraction of mass in the tilted component (see Methods). Jointly modeling the numerous tracers of the disk warp at various ages and Galactic radii is the next step in uncovering the distribution of dark matter in the Galaxy.

\newpage

\section*{Methods}

\subsection*{Simulation details \newline}
Here we describe the time-dependent Galactic potential in which we compute the orbits of tracer particles. The potential is a summation of the following components. The halo has a total mass of $8\times10^{11} \text{M}_\odot$, 70\% of which is in a spherical halo with NFW\cite{NFW97} radial profile with scale radius $r_s=15\text{ kpc}$, and 30\% of which is in a tilted, triaxial halo. The triaxiality is 10:8.1:7.3\cite{han22b} and the tilt angle is $25^\circ$\cite{han22b}. The radial density along the principal axes of the triaxial halo follows an NFW profile with scale radius $r_s=30\text{ kpc}$\cite{han22a}. The disk has two components: a ``thick'' disk with fixed mass at $6\times10^{9} \text{M}_\odot$, scale radius 2 kpc, and scale height 0.9 kpc\cite{bland-hawthorn16}, and a ``thin'' disk that linearly increases in mass up to $3.5\times10^{10} \text{M}_\odot$ at present day, with fixed scale radius 2.6 kpc and scale height 0.3 kpc\cite{bland-hawthorn16} assuming a Miyamoto-Nagai potential\cite{miyamoto75}. The initial mass of the thin disk is $1.3\times10^{10} \text{M}_\odot$. Lastly, we include a spherical bulge with Hernquist radial profile\cite{Hernquist90} and fixed mass $1.8\times10^{10} \text{M}_\odot$ and scale radius of 1 kpc.

The tracer particles are initialized on circular orbits at radii sampled from a truncated exponential distribution, discarding any stars sampled within 1 kpc or beyond eight times the scale radius from the Galactic center. The scale radius increases linearly with time to 8 kpc at present day. The final scale radius is chosen to be where the $\text{H}_2$ mass surface density\cite{heyer15} is $1/e$ of the maximum value. At each time step of the simulation, we spawn new tracer particles at a rate that is commensurate with the growth of the disk, culminating in 50,000 star particles and 50,000 gas particles. We calculate orbits using the \texttt{gala} python package\cite{gala:joss} using a fixed 1 Myr timestep and standard Leapfrog integrator.

While the star particles are collisionless, the gas particles are collisional and follow an inelastic scattering prescription. If a gas particle comes within $0.1 \text{ kpc}$ of another gas particle and they have negative relative speeds, they exchange relative velocities and lose $10\%$ of the collective kinetic energy. This method of simulating gas particles is valid when the velocity dispersion is low\cite{negroponte83,carlberg85}, as it is the case for circular orbits. We note that all of our disk particles remain on circular orbits throughout all 5 Gyr, with the most eccentric orbits exhibiting less than 0.1\% of their angular momentum off of the Galactic $Z$ axis. Lastly, at each timestep, we allow for a probabilistic change of radius of each particle commensurate to what is measured for the Galaxy\cite{frankel18,sellwood2002}.

\subsection*{Warp Modeling \newline}
In Figure \ref{fig:lit} we plot a fit to the warp in the simulation using an analytical formula that is a power-law in radius and a sinusoid in azimuth:
\begin{align*}
    Z(R \geq R_w) &= A\times (R - R_w)^{b} \times \sin{(\phi - \phi_w)} \\
    Z(R<R_w) &= 0
\end{align*}
Here, $R$ and $Z$ are Galactocentric cylindrical coordinates, $A$ is the amplitude of the warp, $b$ the power-law index, $\phi_w$ the orientation of the warp, and $R_w$ the onset radius of the warp. This function has also been used to fit the Cepheids data\cite{chen19}. The fit is performed using a maximum likelihood method.

\subsection*{Which Stars Comprise the Warp? \newline}

In Extended Data Figure \ref{fig:ED_1} we color the star particles by their birth radius. We find a clear correlation in the birth radius of the star and its final warp amplitude. We can thus understand the prominence of the warp in young stars as a result of the ``inside-out'' growth of the disk\cite{frankel19}, in which the birth radius of a particle correlates inversely with age. Since young stars can be born at larger radii than old stars, they trace a cleaner and larger warp. Due to radial migration\cite{sellwood2002}, old stars that are born in the inner Galaxy can also migrate outwards to eventually trace the warp. This mechanism can explain why older stars appear to have smaller warp amplitudes in observations\cite{wang20}.

\subsection*{Timescale of the Warp \newline}

In Extended Data Figure \ref{fig:ED_2} we show the time evolution of the warp amplitude at a fixed radius $R=16\text{ kpc}$. The error bars indicate $1\sigma$ statistical uncertainties in the warp fit. At $t=0$, the disk is initialized to have no warp. Within the first few hundred Myr, the warp reaches its maximum amplitude. This is consistent with the rotation period of the disk at 16 kpc, which is approximately $400\text{ Myr}$. Thus, this Figure shows that the disk responds to the tilted dark halo quickly, within one rotation period of the disk. The warp amplitude experiences a transient oscillatory phase until $1500\text{ Myr}$, then converges to a steady state. This transient phase is likely a numerical effect, since there are not many stars out at $R=16\text{ kpc}$ at these times (the tracer particle scale length is 3 kpc at $t=0$ and 4.5 kpc at $t=1500\text{ Myr}$) and the warp amplitude is determined by a small number of particles.

\subsection*{Variations to Model Parameters \newline}

In this study, we have intentionally avoided ``fitting'' model parameters to the data, in order to demonstrate that no tuning is required for a tilted dark halo to reproduce the observed warp/flare. In Extended Data Figures \ref{fig:ED_3} and \ref{fig:ED_4} we show how changes in the model parameters can affect the warp. In the former, we vary (1) the scale length of the tilted dark halo and (2) the present-day scale-length of the tracer particles. In the latter, we vary (1) the tilt angle of the dark halo and (2) the mass fraction of the tilted component of the dark halo compared to the spherical component. Aside from the parameters being modulated, all other parameters are fixed to their values from the simulation presented in Figure \ref{fig:model age}. In each panel, we calculate the fraction of stars that are off the $Z=0$ plane by more than 0.25 kpc, $N_\text{warp} / N_\text{plane}$. If $N_\text{warp} / N_\text{plane}$ is greater than $0.5\%$, we fit a warp model (dotted lines) and show the warp amplitude at 20 kpc as $Z_{20}$. The disk warps in all models except for the case of a non-tilted halo (last row of Extended Data Figure \ref{fig:ED_4}). These Figures demonstrate that warping is a general response of the disk to a tilted dark halo. Furthermore, a more complete observational picture of the Galactic warp can help constrain properties of the tilted dark matter halo, such as its mass fraction and tilt angle.

\subsection*{Context in Cosmological Simulations \newline}

A key assumption of our idealized simulation is the fixed orientation of the halo with respect to the growing disk. It is thus important to understand to what extent these assumptions are applicable in a simulation with a live halo and self-consistently growing disk. Specifically, the time that it takes for a tilted dark halo to eventually align with the disk is an open question. If this timescale is substantially longer than the warp onset timescale (a few hundred Myr, see Extended Data Fig. \ref{fig:ED_2}), then the mechanisms studied in our idealized simulation can apply more generally to slowly-changing halos.

To address this question, Han et al. (in prep) analyze Milky Way-like galaxies in the Illustris TNG50 cosmological magneto-hydrodynamic simulations\cite{pillepich19, nelson19}. They first show that a significant fraction of MW analogs in TNG50 have present-day tilted dark halos at $r<50\text{ kpc}$. About 50\% of halos are tilted more than $10^\circ$, and 25\% of halos are tilted more than $20^\circ$. Then, they identify a galaxy that experiences a major merger at $\sim7\text{ Gyr}$ ago that results in a tilted dark halo. For this galaxy, the angle of misalignment of the disk and the dark halo reduces from $50^\circ$ to $20^\circ$ over 5 Gyr. Furthermore, the disk of this galaxy warps shortly after the onset of the tilted dark halo with a delay time of $<1\text{ Gyr}$. The galaxy does not have any massive satellites at the time of the onset of the warp. TNG50 was not tuned in any way to produce these results; rather, the long-lived tilted dark halos and warps emerge naturally in the simulations. These results show that tilted dark halos can (1) be long-lived in a live, cosmological environment, and (2) generate warps in galactic disks on timescales shorter than the change in the tilt angle. Furthermore, this result shows that the Milky Way's dark halo was likely more tilted in the past and has decreased to its current value ($\sim25^\circ$) at present day.

\bibliographystyle{naturemag}
\bibliography{refs}

\setcounter{figure}{0}
\makeatletter 
\renewcommand{\figurename}{Extended Data Fig.}
\makeatother

\begin{figure*}[t!]
    \centering
    \includegraphics[width=0.5\textwidth]{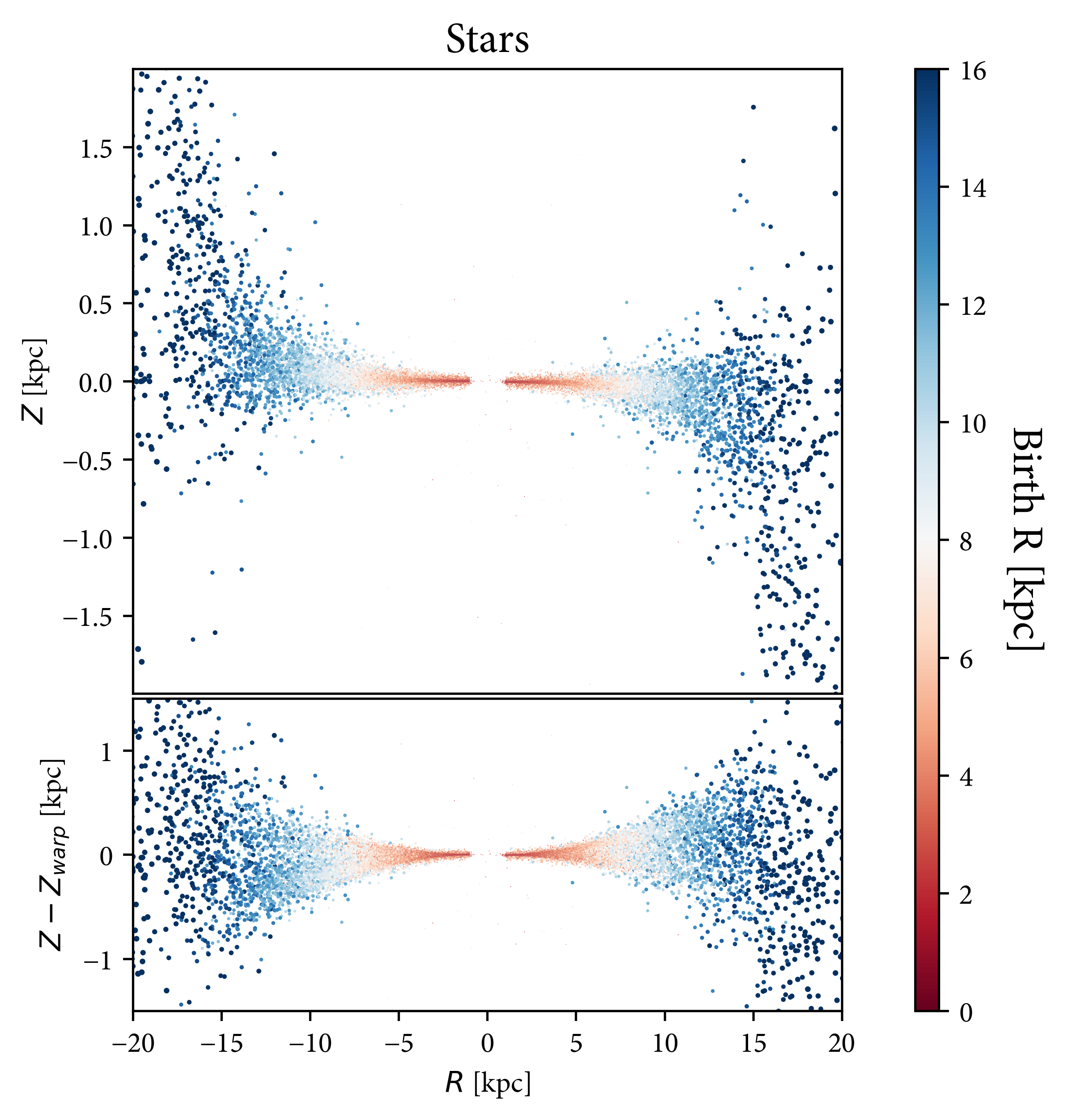}
    \caption{Present-day distribution of star particles in the simulation in Galactocentric cylindrical coordinates. Negative R indicates azimuthal angles that are within $90^\circ$ of the Northern warp, and positive R indicates azimuthal angles that are within $90^\circ$ of the Southern warp. The top panels show the warp, and the bottom panels show the vertical deviation from the average warp. The vertical deviation systematically increases toward the outer Galaxy, demonstrating the flare of the disk. Points that are in the outer Galaxy are plotted with larger circles. Particles colored by birth radius reveal that the magnitude of the warp correlates strongly with the birth radius.}
    \label{fig:ED_1}
\end{figure*}

\begin{figure*}[h]
    \centering
    \includegraphics[width=0.7\textwidth]{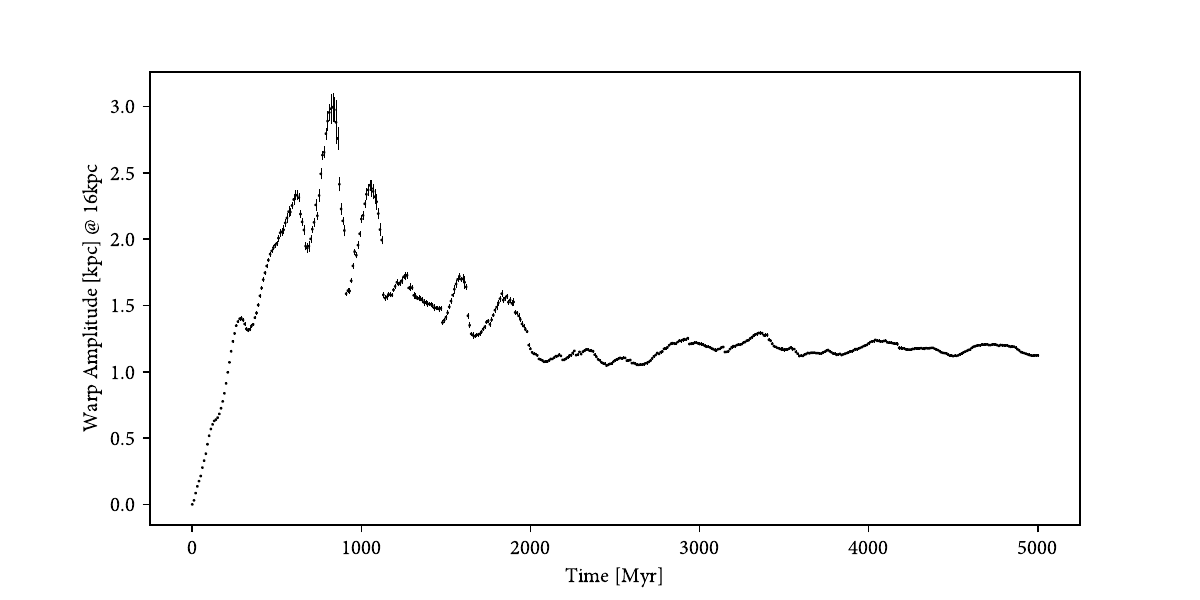}
    \caption{
    Time evolution of the warp amplitude at fixed radius $R=16\text{ kpc}$. At $t=0$, the disk is initialized to have no warp. Error bars indicate $1\sigma$ statistical uncertainty in the warp fit. Within a few hundred Myr, the warp reaches maximum amplitude. After a transient oscillatory phase from $t=500-1500\text{ Myr}$, the warp reaches a steady-state amplitude. This plot demonstrates that the warp onsets quickly, within one rotation period of the disk ($400\text{ Myr}$ for a star at 16 kpc).
    }
    \label{fig:ED_2}
\end{figure*}

\begin{figure*}[t!]
    \centering
    \includegraphics[width=\textwidth]{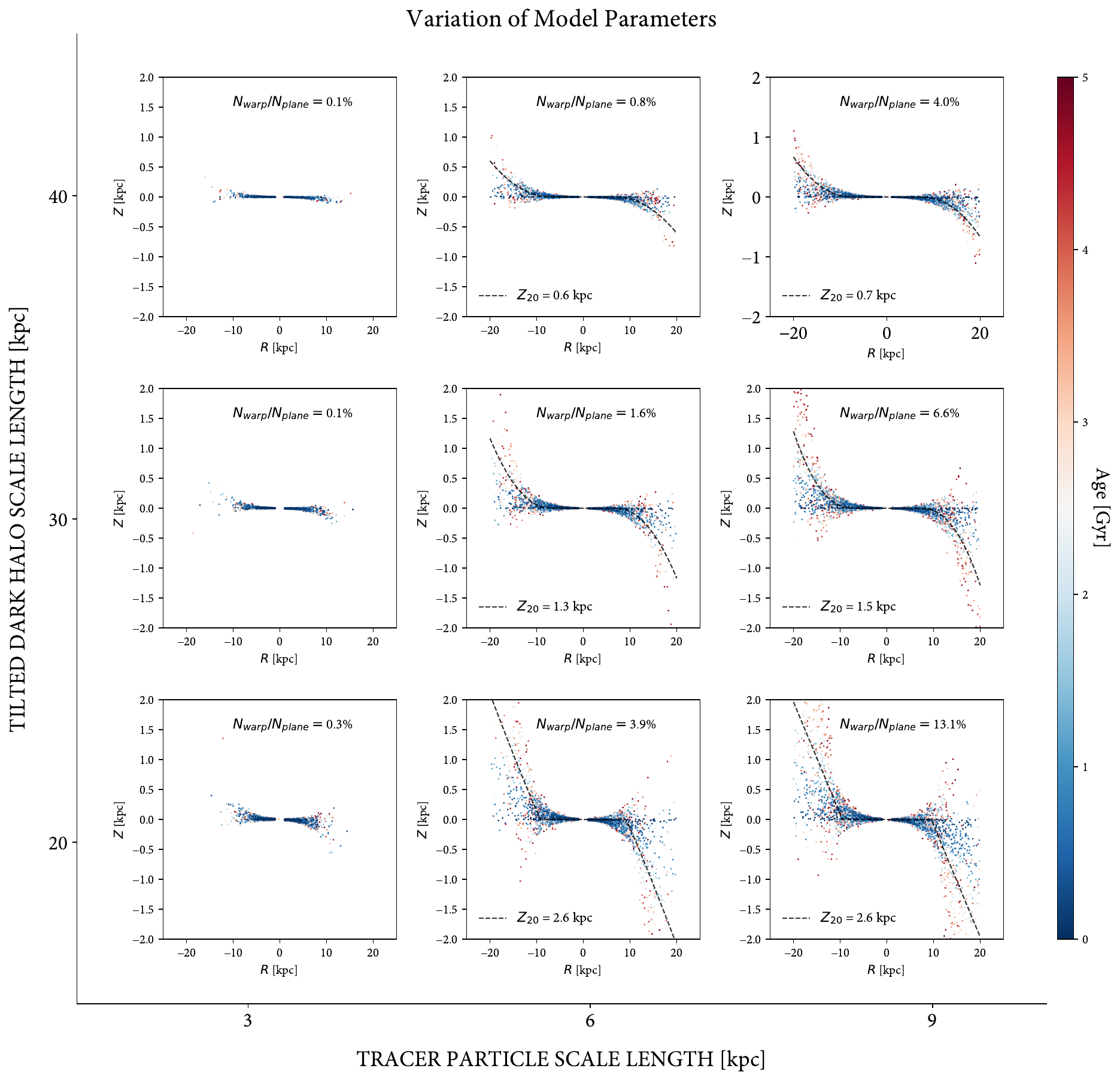}
    \caption{
    Variation of (1) the scale length of the tilted component of the dark halo and (2) the present-day scale length of the tracer particles. Aside from the two parameters being modulated, all other parameters are fixed to the simulation presented in Figure \ref{fig:model age}. In each panel, we show the fraction of stars that are off the $Z=0$ plane by more than 0.25 kpc, $N_\text{warp} / N_\text{plane}$. If this fraction is greater than $0.5\%$, we fit a warp model (dotted lines) and show the warp amplitude at 20 kpc as $Z_{20}$. The warp amplitude correlates positively with the scale length of the disk, and anti-correlates with the scale length of the tilted dark halo. 
    }
    \label{fig:ED_3}
\end{figure*}

\begin{figure*}[t!]
    \centering
    \includegraphics[width=\textwidth]{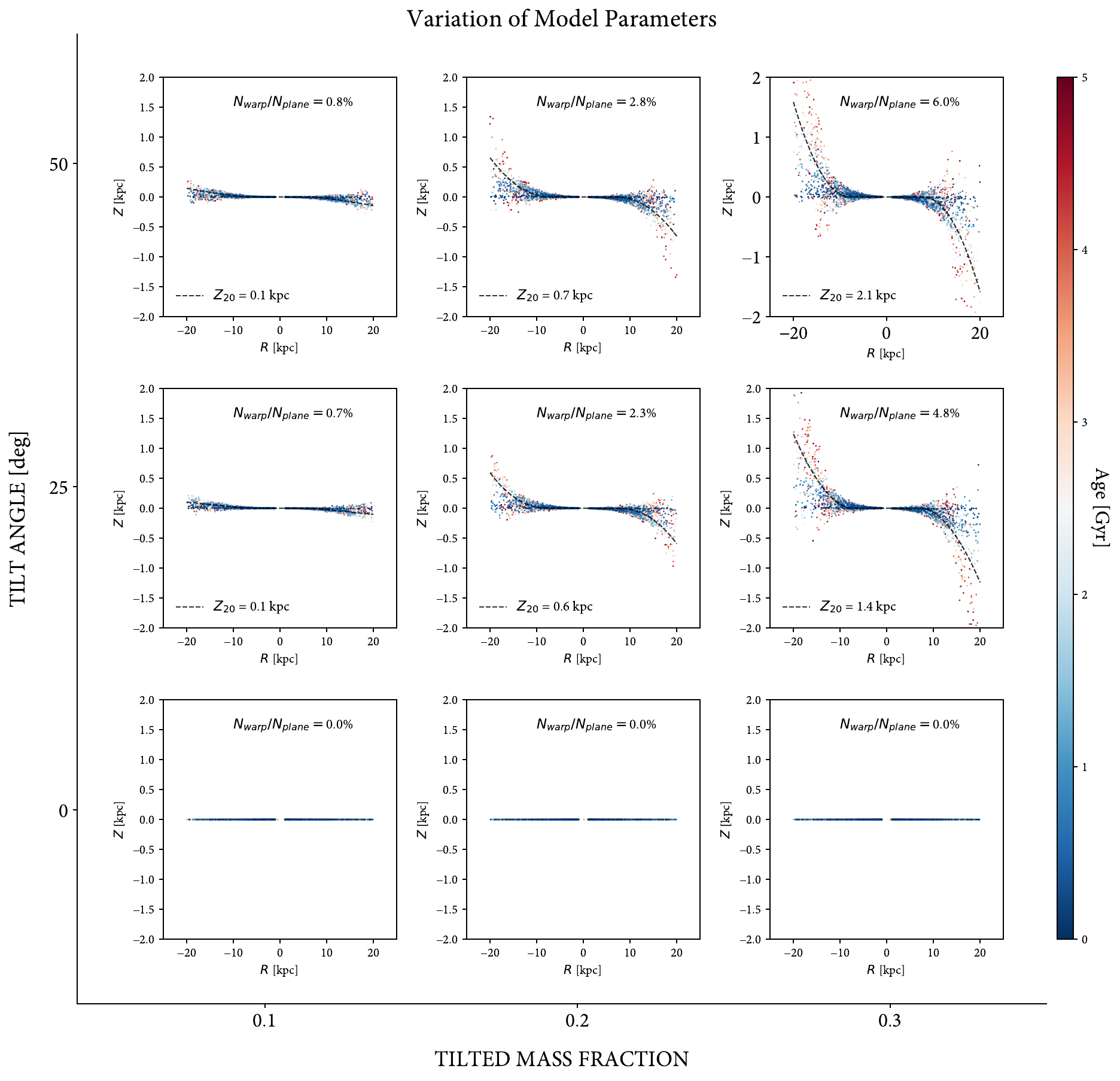}
    \caption{
    Variation of (1) the tilt angle of the dark halo (2) the mass fraction of the tilted component of the dark halo. Aside from the two parameters being modulated, all other parameters are fixed to the simulation presented in Figure \ref{fig:model age}. Similar to Extended Data Figure \ref{fig:ED_3}, we show the fraction of stars that are off the $Z=0$ plane by more than 0.25 kpc, $N_\text{warp} / N_\text{plane}$. If this fraction is greater than $0.5\%$, we fit a warp model (dotted lines) and show the warp amplitude at 20 kpc as $Z_{20}$. The warp amplitude correlates positively with both the tilt angle and the mass fraction of the tilted halo. 
    }
    \label{fig:ED_4}
\end{figure*}

\end{document}